# A theory of chemical Reactions in biomolecules in solution: generalized Langevin mode analysis (GLMA)

Fumio Hirata, Institute for Molecular Science*

*Professor Emeritus

**ABSTRACT**

The generalized Langevin mode analysis (GLMA) is applied to chemical reactions in biomolecules in solution. The theory sees a chemical reaction in solution as a barrier crossing process, similar to the Marcus theory. The barrier is defined as the crossing point of two free-energy surfaces which are attributed to the reactant and product of the reaction. It is assumed that the both free-energy surfaces are *quadratic* or *harmonic*. The assumption is based on the Kim-Hirata theory of structural fluctuation of protein, which proves that the fluctuation around an equilibrium structure is *quadratic* with respect to the structure or atomic coordinates. The quadratic surface is a composite of many harmonic functions with different modes or frequencies. The height of the activation barrier will be dependent on the mode or frequency, less the frequency, lower the barrier. So, it is essential to decouple the fluctuational mode into a hierarchical order. GLMA is impeccable for this purpose.

It is essential for a theoretical study of chemical reactions to chose a *reaction coordinate* along which the reaction proceeds. We suppose that the mode whose center of coordinate and/or the frequency changes most before and after the reaction is the one relevant to the chemical reaction, and choose the coordinate as the reaction coordinate. The rate of reaction along the reaction coordinate is $k_{rate} = \nu \exp\left[-\Delta F^{(\dagger)} / k_B T\right]$, which is similar to the Marcus expression for the electron transfer reaction. In the equation, $\Delta F^{(\dagger)}$ is the activation barrier defined by $\Delta F^{(\dagger)} \equiv F^{(r)}\left(Q^{\dagger}\right) - F^{(r)}(Q_{eq}^{(r)})$, where $F^{(r)}(Q_{eq}^{(r)})$ and $F^{(r)}\left(Q^{\dagger}\right)$ denote the free energies at equilibrium $Q_{eq}^{(r)}$, and the crossing point $Q^{\dagger}$, respectively, both on the free energy surface of the reactant.

## I. Introduction

A life is maintained by a variety of chemical reactions in solution, such as photo synthesis,

enzymatic reactions, and so forth.[1] The time constant of those reactions ranges from nanoseconds to minutes or hours in some cases. Why the time constant of the reactions is so widely spread is because the reaction is always associated with a reorganization of the molecular structure, or *isomerization*, of proteins in which the reaction is taking place. A reaction will not be completed until the structure of biomolecules and its environment, or *aqueous solution*, are equilibrated *thermodynamically* to the product state of the reaction. Here, "thermodynamically" means that the product state should be at a minimum of the *free energy* surface including the *solvation* free energy, not just the *potential energy* of a *naked* biomolecule.[2]

Although the time constant of the chemical reactions is widely spread depending on the reaction, they are well separated in many cases of interest. Let us sketch few examples of such cases in the following.

So-called *protein folding* is an *isomerization* reaction in which the many dihedral-angles around chemical bonds constructing the backbone structure should change harmoniously from gauche to trans or vice versa. It is a global change of the structure. On the other hand, the global change of structure is associated with many *local isomerization reactions* with a variety of time constants.[3-5]

Let us see the separation of the time constant in the enzymatic reactions. An enzymatic reaction consists essentially two *chemical processes* as been clarified by Michaelis and Menten: binding or unbinding of a substrate at a cavity of protein (enzyme), and a *chemical reaction* or recombination of atoms in substrate molecules inside the cavity.[6] In general, time constants of those processes are well separated. In fact, it is the basic assumption in the Michaelis-Menten theory that the reaction between bound and unbound states, or the *molecular recognition process*, is so quick that they are already in *equilibrium*. However, the molecular recognition process itself has some time constant which may be short, but it is of current interest.[7,8]

A group of proteins called "molecular channels" has an interesting mechanism, referred to as "gating" which plays a role of *valve*, controlling the binding/unbinding of small ions and molecules to be transported to/from the channel pore. The gating is nothing but a local conformational change, or *isomerization*, of the biomolecule. [9,10]

Another example of the time separation in a chemical process in protein is seen in an energy-transfer or/and charge-transfer reaction, induced by photo excitations. A typical example of such reactions is the twisted intramolecular charge transfer (TICT) reaction, in which the charge transfer is conjugated with a local conformational change, or "twisting", of the molecule. In such a case, the rate of local conformational change largely determines the reaction rate.[11,12]

Purpose of the present paper is to propose a theoretical method to decompose the fluctuation having many different time constants in a hierarchical order. For that purpose, it is convenient to employ the terminology, "mode", to express each fluctuation with different time constant. In fact, the terminology is the one which has been used not only in the conventional theories of the chemical reaction, but also in the normal mode analysis (NMA) of protein. [13] Using the NMA, the entire fluctuation of protein composed of many fluctuations with different modes is decoupled into many single modes, each having different frequency or time constant, which have a hierarchical structure. Unfortunately, earlier treatments of protein with NMA have stayed just in a conceptual proposal, since the protein they have handled was that in *vacuum*, not a *realistic* one in aqueous solutions.[14-17] More realistic treatments have been made by Kitao et. al. based on the molecular dynamics (MD) simulation. [18,19] The authors have applied the principal axis analysis, or the principal component analysis (PCA), to a trajectory obtained from MD simulations for a protein in water to decouple the fluctuational modes of the protein. Later, PCA was applied to the protein folding by Maisuradae et. al. [20] The PCA is essentially equivalent to NMA, that is, a method to diagonalize the coordinate axes of the *potential energy* surface of a system consisting of a protein and water molecules.

Recently, we have proposed a new method to decompose the fluctuation of a biomolecule in solvent into independent modes with different frequencies in a hierarchical order, based on the combined generalized Langevin and RISM/3D-RISM theories proposed by Kim and Hirata.[21-23] The method is referred to as Generalized Langevin Mode Analysis (GLMA). The most important difference between PCA and GLMA is whether the *solvation free energy* is taken into consideration or not. The method was applied to an aqueous solution of a dipeptide to calculate the RIKES spectrum measured by Giraud and Klaas.[24] In spite that there are small mismatches between the systems measured by experiment and that calculated by the theory, the theoretical result has reproduced the observed spectrum reasonably well, except for the differences caused by the mismatches.

In the present paper, we propose a new theory to calculate the rate constant of chemical reactions taking place in biomolecules in solution, based on the GLMA theory. In the new theory, a chemical reaction in solution is regarded as a "barrier crossing" problem, similar to the Marcus theory of the electron transfer reaction.[25-30] The barrier is defined as a crossing point of two free-energy surfaces, each of which is attributed to the reactant and product of the reaction. The most essential premise of the theory is that the both free-energy surfaces, reactant and product, are "quadratic" or "harmonic". The premise concerning the free energy surface is based on the generalized Langevin equation for structural fluctuation of protein derived by Kim and Hirata, according to which the force acting on each atom in the equation is linear to their displacement from the equilibrium state. [22,23] The quadratic surface is a composite of many harmonic functions with different modes or frequencies, which are hierarchically ordered. The height of the activation barrier will be also ordered hierarchically: less the frequency, higher the barrier. So, the reaction rate should be analyzed according to the mode of fluctuation. It is the fundamental concept on which the present theory is based.

## II. Theory
### II-1. Generalized Langevin theory of a protein in solvent

The treatment here is almost identical to that proposed by Lamm and Szabo except for the model and the starting equation.[31] Lamm and Szabo started from the classical Langevin equation applied to a connected harmonic-oscillator, proposed while ago by Wang and Uhlenbeck, in which the atomic motion evolves on a *potential-energy* surface produced by the *direct interactions* among atoms. [31,32] The theory proposed here starts from the *generalized* Langevin equations of the structural fluctuation of a protein in solution, derived by Kim and Hirata. In the dynamics, each atom is driven by a force derived from the *free-energy* surface, that includes not only the *direct interaction* among atoms but also the solvent induced force. [22,23] A brief review of the theory is given as follows.

The structural fluctuation of a biomolecule in solution is defined by,

$$\Delta \mathbf{R}_\alpha(t) \equiv \mathbf{R}_\alpha(t) - \langle \mathbf{R}_\alpha \rangle \tag{1}$$

where $\mathbf{R}_\alpha(t)$ is the current position of atom $\alpha$, and $\langle \mathbf{R}_\alpha \rangle$ defines the equilibrium position of the atom. The generalized Langevin equations of the structural fluctuation are written as,

$$\frac{d\Delta \mathbf{R}_\alpha(t)}{dt} = \Delta \mathbf{V}_\alpha(t) \tag{2}$$

$$\frac{d\Delta \mathbf{V}_\alpha(t)}{dt} = -\frac{k_B T}{M_\alpha} \sum_\beta \left(\mathbf{L}^{-1}\right)_{\alpha\beta} \cdot \Delta \mathbf{R}_\alpha(t)$$
$$- \int_0^t ds \sum_\beta \frac{1}{M_\alpha} \Gamma_{\alpha\beta}(t-s) \cdot \Delta \mathbf{V}_\beta(s) + \mathbf{W}_\alpha(t) \tag{3}$$

where **L** is the variance-covariance matrix of structural fluctuation, defined by

$$L_{\alpha\beta} \equiv \langle \Delta \mathbf{R} \Delta \mathbf{R} \rangle_{\alpha\beta}. \tag{4}$$

$\Gamma_{\alpha\beta}(t)$ is the friction kernel, and $\mathbf{W}_\alpha(t)$ represents the random force. The variance covariance matrix is related to the second derivative of the free energy surface $F(\{\mathbf{R}\})$ with respect to the atomic coordinates of protein as,

$$k_B T (\mathbf{L}^{-1})_{\alpha\beta} = \frac{\partial^2 F(\{\mathbf{R}\})}{\partial \mathbf{R}_\alpha \partial \mathbf{R}_\beta}. \tag{5}$$

The free energy surface is defined as the sum of the conformational energy $U(\{\mathbf{R}\})$ of the solute and the solvation free-energy $\mu(\{\mathbf{R}\})$ as follows,

$$F(\{\mathbf{R}\}) = U(\{\mathbf{R}\}) + \mu(\{\mathbf{R}\}). \tag{6}$$

Multiplying Eqs. (1) and (2) by $\Delta \mathbf{R}_\gamma(0)$, and taking the ensemble average, one finds

$$\frac{d\langle \Delta \mathbf{R}_\alpha(t) \Delta \mathbf{R}_\gamma(0) \rangle}{dt} = \langle \Delta \mathbf{V}_\alpha(t) \Delta \mathbf{R}_\gamma(0) \rangle, \tag{7}$$

$$\begin{aligned}\frac{d\langle \Delta \mathbf{V}_\alpha(t) \Delta \mathbf{R}_\gamma(0) \rangle}{dt} &= -\frac{k_B T}{M_\alpha} \sum_\beta (\mathbf{L}^{-1})_{\alpha\beta} \cdot \langle \Delta \mathbf{R}_\alpha(t) \Delta \mathbf{R}_\gamma(0) \rangle \\ &\quad - \int_0^t ds \sum_\beta \frac{1}{M_\alpha} \Gamma_{\alpha\beta}(t-s) \cdot \langle \Delta \mathbf{V}_\beta(s) \Delta \mathbf{R}_\gamma(0) \rangle \\ &\quad + \langle \Delta \mathbf{W}_\alpha(s) \Delta \mathbf{R}_\gamma(0) \rangle \end{aligned} \tag{8}$$

The last term in the second equation vanishes by definition, because $\Delta \mathbf{R}_\gamma(0)$ is one of the dynamic variables, which is orthogonal to the random force $\Delta \mathbf{W}_\alpha$, that is,

$$\langle \Delta \mathbf{W}_\alpha(s) \Delta \mathbf{R}_\gamma(0) \rangle = 0. \tag{9}$$

We further assume that the friction coefficient is local in time, or

$$\Gamma_{\alpha\beta}(t-s) = \zeta_{\alpha\beta} \delta(t-s). \tag{10}$$

Taking those into account, the second equation becomes,

$$\frac{d\langle\Delta\mathbf{V}_\alpha(t)\Delta\mathbf{R}_\gamma(0)\rangle}{dt} = -\frac{k_B T}{M_\alpha}\sum_\beta K_{\alpha\beta}\cdot\langle\Delta\mathbf{R}_\beta(t)\Delta\mathbf{R}_\gamma(0)\rangle \\ -\sum_\beta \zeta_{\alpha\beta}\langle\Delta\mathbf{V}_\beta(s)\Delta\mathbf{R}_\gamma(0)\rangle \tag{11}$$

where $K_{\alpha\beta}$ and $\zeta_{\alpha\beta}$ are defined by

$$K_{\alpha\beta} \equiv \frac{k_B T}{M_\alpha}\left(\mathbf{L}^{-1}\right)_{\alpha\beta}, \quad \zeta_{\alpha\beta} \equiv \frac{\Gamma_{\alpha\beta}}{M_\alpha}. \tag{12}$$

Eqs. (7) and (11) are written in a matrix form as

$$\frac{d}{dt}\begin{pmatrix}\mathbf{C}(t)\\ \dot{\mathbf{C}}(t)\end{pmatrix} = \mathbf{A}\begin{pmatrix}\mathbf{C}(t)\\ \dot{\mathbf{C}}(t)\end{pmatrix}, \tag{13}$$

where C(t) and $\dot{\mathbf{C}}(t)$ are defined by

$$C_{\alpha\gamma}(t) \equiv \langle\Delta\mathbf{R}_\alpha(t)\Delta\mathbf{R}_\gamma(0)\rangle, \quad \dot{C}_{\alpha\gamma}(t) \equiv \langle\Delta\mathbf{V}_\alpha(t)\Delta\mathbf{R}_\gamma(0)\rangle, \tag{14}$$

and the matrix **A** is defined by the following equation.

$$\mathbf{A} = \begin{pmatrix} 0 & 1 \\ -\mathbf{K} & -\zeta \end{pmatrix}. \tag{15}$$

We refer to the matrix A as "generalized force constant (GFC) matrix". As it is clear from the definition, GFC carries not only the information concerning the second derivative of the *free energy* surface, but also the influence due to the *friction* from solvent. So, the generalization of the force constant compared to the usual force constant of a molecule situated in *vacuum* is double fold: inclusion of the solvation free-energy and the friction from solvent. It should be noted, however, that the basic functional form of the free energy surface on which the structure fluctuates is quadratic. The kinetic energy is also *quadratic* function of the momentum. Therefore, the harmonic analysis similar to the normal mode analysis is valid for the generalized *Hessian* (Eq. 15) in the phase space.

Unfortunately, the friction coefficient acting on each atom in a biomolecule is not well clarified at the moment. It is because the solvent environment around each atom in a biomolecule such as protein is extremely inhomogeneous: atoms on the surface of protein is exposed well to solvent, while those buried inside the molecule are not so. Therefore, in the present work, we just focus our attention on the free energy surface $F(\{\mathbf{R}\})$ in the configurational space, defined by Eq. (6), and on the Hessian matrix $K_{\alpha\beta}$ defined by Eqs. (5) and (12).

## II-2. Generalized Langevin Mode analysis (GLMA) of chemical reaction in solution

It should be noted that the free energy surface in the cartesian-coordinate space described in the previous section is a composite of many surfaces with different curvature, or *mode*, which has a hierarchical structure. It will be rational to think that the rate of a chemical reaction taking place in a macromolecule such as protein is highly dependent on the *mode* of structural fluctuation.

As an example, let's think of the folding reaction of a protein molecule. Imagine a prototypical folding-reaction, or a two-states transition from denatured to native states. The free energy surface of such a *reaction* is represented by two curves with wells along a reaction coordinate, each of which is expressed by a quadratic function. The crossing point of the two curves is identified as the *transition state* of the reaction. Apparently, the height of the activation barrier essentially determines the rate of reaction. It should be noted however that the overall reaction-rate of folding is a composite of many *chemical-reactions*, for examples, gauche-trans isomerization around dihedral angles, having different timescales hierarchically ordered depending on their modes. Such *chemical reactions* with different modes may also be of interest for exploring the rate of reactions taking place in protein, such as in an enzyme.

It is the advantage of the Generalized Langevin mode analysis (GLMA) to decompose the mode, and to project the free energy surface onto a coordinate axis of a single mode. Such decoupling of modes can be made by diagonalizing the Hessian matrix $K_{\alpha\beta}$ defined by Eq. (12). The diagonalization is carried out by a standard linear-algebra in the multi-dimensional space.

First of all, we must find the origin of the coordinate space, at which the free energy of the molecule becomes minimum in the configuration space spanned by the 3*N*-6 atomic coordinates. The origin can be found by translating the origin of the configuration space to the center of the free energy surface. Alternatively, the origin can be found by minimizing the free energy surface, where the two origins should coincide by definition.

Then, the axis is *rotated* to get the new axis which diagonalizes the Hessian matrix.

The free energy projected onto the new coordinate-space, or the *normal-mode* coordinates space, can be expressed in the standard form as

$$F(\{Q\}) = \sum_{m}^{3N-6} K_m (Q_m - Q_{m,eq})^2 \tag{15}$$

where $Q_m$ denotes the normal-mode coordinates of the *m*-th mode, $Q_{m,eq}$ is the coordinate at the origin, or the equilibrium structure, and $K_m$ represents the Hessian or the "force constant" associated with the *m*-th mode. It will be worthwhile to note here that our analysis of the chemical-reaction rate requires information of the equilibrium structures of the molecule before and after the reaction. (In that respect, the problem is entirely different from a prediction of a molecular structure, such as the "protein folding", in which the essential task is to find an equilibrium structure in the given thermodynamic condition.)

## II-3. GLMA for the rate of chemical reactions in solution

Eq. (15) is applied to formulate the rate of chemical reaction in terms of the activation-barrier crossing. First of all, we have to define the mode which is relevant to the chemical reaction. We put the following hypothesis. *The mode whose center of coordinate and/or the frequency changes most before and after the reaction is the one relevant to the chemical reaction, and choose the coordinate as the reaction coordinate.* For examples, the coordinates corresponding to the least eigen-value is likely to be the reaction coordinate in case of the protein folding.

Let us denote the two free energy surfaces for reactant and product, projected on to the reaction coordinate by,

$$F^{(r)}(Q) = K^{(r)}(Q - Q_{eq}^{(r)})^2 + F_{eq}^{(r)} \tag{16}$$

$$F^{(p)}(Q) = K^{(p)}(Q - Q_{eq}^{(p)})^2 + F_{eq}^{(r)} \tag{17}$$

where the suffices "*r*" and "*p*" specify the quantities concerning reactant and product, and $F_{eq}^{(r)}$ in the equations represents the free energy of the equilibrium structure of the reactant system.

The barrier is defined as a crossing point of the two parabolas expressed by Eqs. (16) and (17). It is a trivial mathematical problem to find the crossing point of two simple parabolas like Eqs. (16) and (17). Let's denote the crossing point of the m-*th* normal mode coordinate by $Q^{\dagger}$. Then, the activation barrier from the reactant side will be readily given by the following equation.

$$\Delta F^{(\dagger)} \equiv F^{(r)}\left(Q^{\dagger}\right) - F^{(r)}(Q_{eq}^{(r)}) \qquad (18)$$

The so-called energy gap $\Delta G^0$ and the solvent reorganization energy, which are relevant to the chemical reaction, can be also defined by the following equations,

$$\Delta G^0 \equiv F_{eq}^{(p)} - F_{eq}^{(r)} \qquad (19)$$

$$\lambda = F^{(r)}\left(Q_{eq}^{(p)}\right) - F^{(p)}\left(Q_{eq}^{(r)}\right) \qquad (20)$$

Finally, the rate of reaction is given by

$$k_{rate} = \nu \exp\left[-\frac{\Delta F^{(\dagger)}}{k_B T}\right] \qquad (21)$$

where $\Delta F^{(\dagger)}$ is expressed in terms of the solvent reorganization energy $\lambda$ by

$$\Delta F^{(\dagger)} = \frac{\left(\lambda + \Delta G^0\right)^2}{4\lambda} \qquad (22)$$

which is the familiar expression given by R. Marcus for the electron transfer reaction.[25-30] The prefactor $\nu$ included in Eq. (21) represents the factor to determine the reaction rate other than the structural and solvent density fluctuation, such as the electronic reorganization that plays an important role when the non-adiabatic process becomes significant. The details discussion of the process is out of scope in the present treatment.[33]

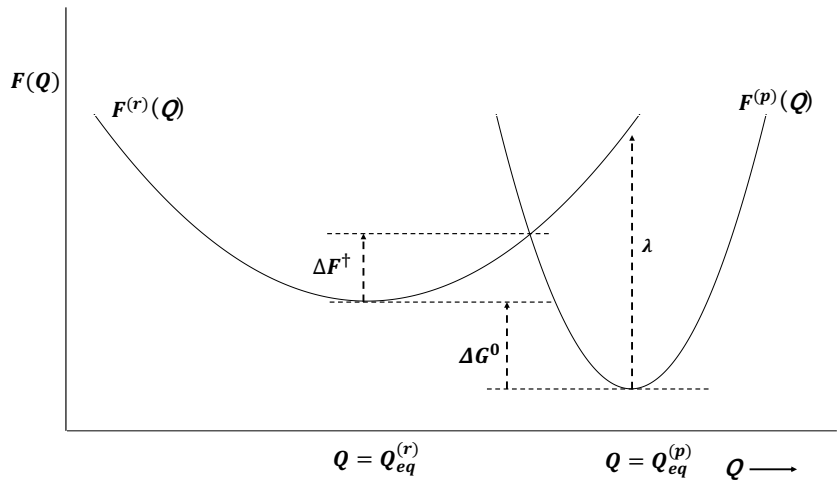

**Fig. 1 Illustration of the free energy surface along the reaction coordinate.**

## III. An illustrative example of possible applications of the theory

It will be worthwhile to provide some image how the theory can be applied to characterize chemical reactions in biomolecules in solution, taking the protein folding as an example. The type of protein folding has been classified by experimentalits roughly into three categories: (a) two- states transition, (b) barrier less transition, and (c) multi-states transition.

The two-state transition is the most popular model in which the unfolded state of protein should cross over a single free-energy barrier in order for folding into the native conformation. The free energy surface along the reaction coordinate in that case is illustrated in **Fig. 2**, which is more or less similar to the case drawn in **Fig. 1**. That is, the parabola corresponding to the unfolded state has the larger variance reflecting the greater entropy, while that corresponding to the folded state has deeper well, implying the greater attractive interactions such as hydrogen-bonds.

The second case is the so-called "barrier-less transition" from unfolded state to the native structure. In that case, the equilibrium structures between the two states should be very different, that is, the minimum points of two parabola should be far distant along the reaction coordinate $Q$. It is also expected that the curvature of the parabola in the unfolded state is so small that activation barrier becomes virtually zero. Such a situation is illustrated in **Fig. 2(b)**. In the figure, the curvature of the parabola $F^{(u)}(Q)$ corresponding to the unfolded state is so small that $\Delta F^{\dagger} / k_B T$ virtually disappears.

Then, what makes the curvature of parabola so small? Let's look back to the equations (4) and (12), which define the curvature or Hessian of the free energy surface. The Hessian matrix is an inverse of the variance-covariance matrix. Therefore, small Hessian means large variance-covariance matrix. Then, which thermodynamic-quantity is relevant to the variance-covariance matrix? That is the *structural entropy* of protein. It suggests that the condition for the barrier-less folding of protein is the denatured state of protein to have large structural entropy.

Such a conclusion concerning the barrier-less folding has been reached already by Chong and Ham who have analyzed the MD trajectory for a protein, provided from ANTON.[34] They have calculated the structural entropy of the protein along the trajectory based on the RISM/3D-RISM theory, and have concluded that the transition is *barrier-less* in accord with the *energy-landscape model* proposed by Wolynes and his coworkers.[35] They have also concluded that the barrier-less transition is essentially related to the large structural-entropy of the denatured state of protein.

The third case often observed in the protein folding is a reaction passing through an intermediate state. For examples, in case of the intermediate state referred to as "molten-globule" state, most of the secondary structures are already formed as in the native conformation. However, packing of those structure into the native conformation is not completed. [36] In such a case, the three-dimensional structures of the reactant or product and the intermediate state may not be so different, thereby the curvatures of the parabola may not be so different as well. Nevertheless, the reaction should pass through an activation barrier, which may not be so high, but not negligible. Such a situation is illustrated in Fig. 2(c).

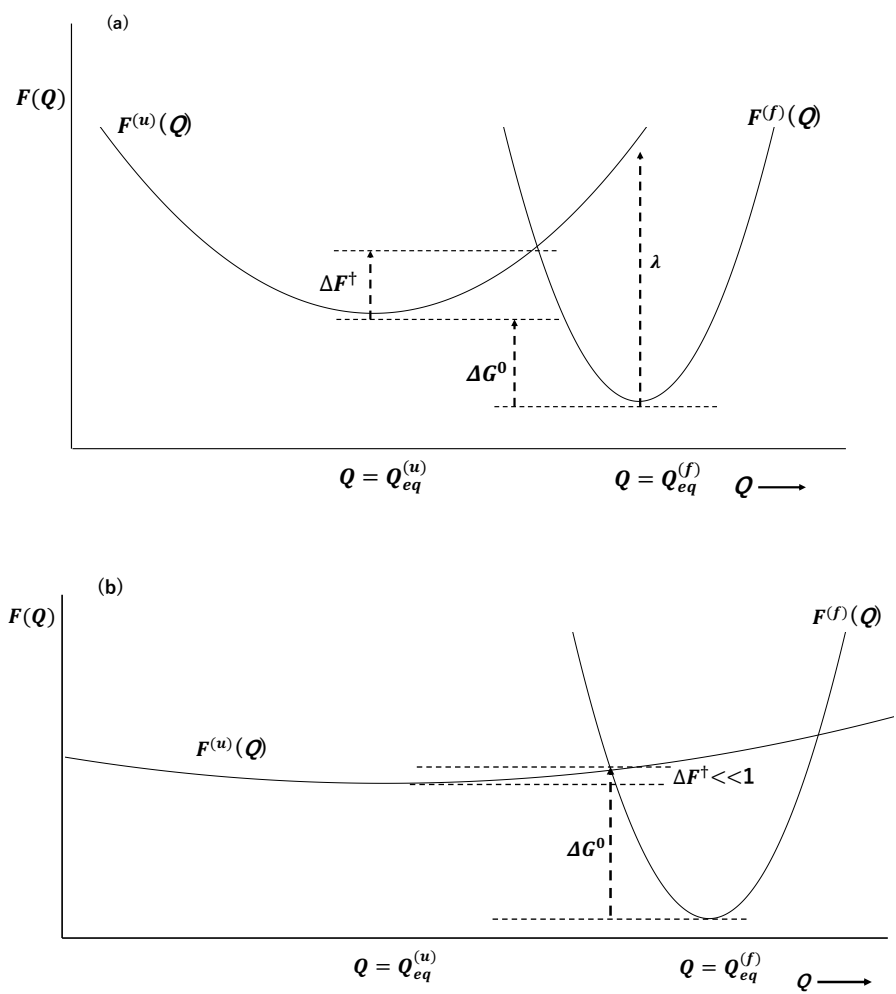

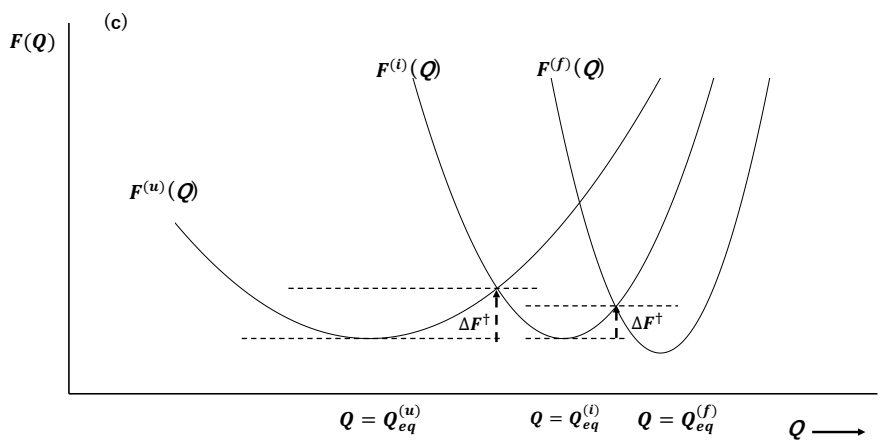

**Fig. 2 Illustration of the three cases of the protein folding-reaction: superscripts "*u*", "*f*" and "*i*" denote "unfolded", "folded", and intermediate states, respectively. (a) the two-states model, (b) the barrier-less folding, (c) the folding through an intermediate state.**

## IV. Discussions

In the present section, we discuss the feasibility of the analysis proposed here to explore the rate of chemical reaction in a macromolecule in solution.

The analysis requires three physical properties to be known as premises: (i) the molecular structure of the reactant and product in terms of atomic coordinates, (ii) the free energy surface and the force constant matrix of the molecules in solution, and (iii) the friction coefficients of atoms in the molecules.

It will be worthwhile to make some comment on availability of those quantities.

### IV-1. Molecular structure of reactant and product

For small biomolecules such as amino-acids or nucleotides in solution, it is a routine work to determine their structures by the experimental technics such as the X-ray crystallography, the electron diffraction, NMR, and so on. If they are sufficiently small, even the quantum chemistry can be employed to determine the structure in gas phase. The structure can be further optimized in solutions using either the molecular dynamics simulation and/or the RISM/3D-RISM methods. On the other hand, it is a non-trivial problem to determine the non-native state of the macromolecules, such as protein and DNA, since it does not have a unique structure by definition. Nevertheless, such structures can be modeled by a random coil state with a gaussian distribution having a rather large variance, around a particular conformation. Such an average structure may be produced by the molecular-dynamics simulation and/or the RISM/3D-RISM in one way or the other.

**IV-2. Free energy surface and force constant matrix**

The free energy surface of the solute is defined by Eq. (6) as a sum of the conformational energy of solute and the solvation free energy.

It is a routine work in the molecular mechanics or dynamics to calculate the conformational energy $U(\{\mathbf{R}\})$ and its second derivatives with respect to the coordinates $\{\mathbf{R}\}$ for a particular conformation of solute. On the other hand, there is no such a simple method to calculate the solvation free-energy, or $\Delta\mu(\{\mathbf{R}\})$, by means of the molecular mechanics and dynamics. It is because the method should sample all the configurational space of solvent for a *single* conformation of the solute, which by itself takes the time equivalent to the Ergordic limit. It is even more difficult and virtually hopeless to calculate the second derivative, or the force constant, of the free energy surface by means of the methods.

Fortunately, such quantities like the solvation free energy and its derivatives can be calculated based on an alternative method of the statistical mechanics, or the RISM/3D-RISM theory. The method to calculate the solvation free energy and its first derivative with respect to atomic coordinates has been proposed while ago, and it has found many applications in the field of biophysics.[2,37,38] The methodology has been well documented, and computer programs to calculate those quantities have been already installed in popular simulation soft wares such as AMBER [39]

It is quite recent that a method to calculate the second derivative of the solvation free energy with respect to the atomic coordinates of solute is proposed.[21] The method employs the idea originally proposed by Yu and Karplus.[40] The idea is to calculate the second order derivative along the course of the iteration procedure to find a converged result for the density distribution function of solvent around a solute. The second derivative itself is regarded as a variable in the iterative scheme for convergence.

**IV-3. Friction coefficient**

In the present article, we have ignored the effect from the friction coefficient onto the reaction rate. However, the effect may become non-trivial in some type of chemical reactions, in which essential parts of molecules in reaction are largely exposed to water. It is a highly non-trivial problem to find the friction coefficient of each atom in a macromolecule in solution by any mean. Here, we propose rather rough but sure way to find the friction coefficient based on the molecular simulation and the 3D-RISM theory.

The source of friction that is proportional to the velocity of a particle is, by definition,

"dissipation" or/and "loss" of momentum and energy of the particle due to collisions and other interactions with particles in environment. In the earlier treatments based on the conventional Langevin theory, a phenomenological expression represented by the Oseen tensor have been employed for the friction matrix [31]. The diagonal elements of the Oseen tensor are the Stokes friction acting on each atom of protein due to solvent, and the off-diagonal elements, referred to as the *hydrodynamic interaction*, describe the force acting on an atom from the other atoms of protein, propagated through solvent. Here, it is worthwhile to emphasize that the both elements of the friction matrix are, by definition, originated from the fluctuation of solvent in the *momentum* space, not that in the *positional* space or the density fluctuation. Suppose that a protein molecule is isolated in vacuum. An atom in the molecule is interacting with the other atoms in the molecule. However, those forces are not considered as "friction," because the momentum and energy of the atom is not *dissipated*, but just *transferred* to the other atoms or vibrational modes. Such energy and momentum transfer associated with the intra-molecular interaction is completely taken care of by the direct interaction terms in our equation. The friction should be exerted on each individual atom of protein, and it should be different from atom to atom depending on the position where the atom is located in the protein. We emphasize again that the friction on each atom of protein is originated entirely from solvent molecules.

We also emphasize that the hydrodynamic interaction should be distinguished from that originated from the potential of mean force, although both are solvent-mediated forces. Actually, in our theory, the two contributions are clearly separated. In Eq. (11), the first term is related to the forces originated from the potential of mean force, while the hydrodynamic interactions are included in the friction-kernel matrix as the off-diagonal elements. Taking the separation of solvent-mediated forces into consideration, we assume that the hydrodynamic interaction between atoms in protein can be negligibly small. It is because the momentum fluctuation of water molecules around an atom, caused by the motion of the atom, is considered to be "thermalized" or to dissipate quickly before the force is propagated to the other atoms in protein. Based on this assumption, we focus our attention on the diagonal terms of the friction matrix.

Based on the arguments above, we propose an approximation for the friction $\zeta_{\alpha\alpha}$ exerted on each atom of protein as follows [23],

$$\zeta_{\alpha\alpha} = f_\alpha \zeta_{\alpha\alpha,bulk} \qquad (23)$$

in which $\zeta_{\alpha\alpha,bulk}$ is the friction exerted on atom $\alpha$ in *bulk* solvent, and $f_\alpha$ is the fraction of the atom exposed to solvent. We define $f_\alpha$ by the following equation based on the radial distribution function (RDF) of solvent around the atom,

$$f_\alpha = \frac{g_w(\sigma; protein)}{g_w(\sigma; bulk)} \qquad (24)$$

where $g_w(\sigma; bulk)$ is the RDF of water in bulk at the contact separation between the atom and a solvent molecule, and $g_w(\sigma; protein)$ is that corresponding to the atom in protein. If the atom is at surface of protein, $f_\alpha$ is close to unity, because the atom is well exposed to solvent, and $g_w(\sigma; protein)$ will become close to $g_w(\sigma; bulk)$. On the other hand, $g_w(\sigma; protein)$ will become small or zero if the atom is buried inside the protein, since there are no or few water molecules around such an atom. Both $g_w(\sigma; bulk)$ and $g_w(\sigma; protein)$ can be readily calculated based on the 3D-RISM/KH theory by making an appropriate definition for the contact separation $\sigma$ between the protein atom and solvent. [2] A typical choice of $\sigma$ can be the position of the first peak in RDF in the bulk solvent.

It is a non-trivial problem to determine $\zeta_{\alpha\alpha,bulk}$ by experimental means, because an atom in protein has a partial charge in general, which is not the case if an atom is isolated by itself in solution: a charged atom in solution exists just in the form of an "ion" which of course has full charges, such as monovalent and divalent ions. Such an atom with a partial charge in solution is just an "imaginary" atom. Then, how can one estimate the friction of such an imaginary atom in solution? In the following, we propose a recipe to determine the friction of an imaginary atom of solution based on the standard MD simulation for an ion with hypothetical *partial-charges* in solution.

***Scheme to determine the friction coefficient of atoms in protein***

1. Assume that $\zeta_{\alpha\beta}(\alpha \neq \beta) = 0$.

2. Make a table of the friction $\zeta_{\alpha\alpha,bulk}$ exerted on the hypothetical atoms in bulk water using MD. In the course of calculation, the radial distribution function $g_w(r;bulk)$ of water around each hypothetical atom is also produced. (Alternatively, $g_w(r;bulk)$ can be calculated by the RISM theory.)

3. Perform the 3D-RISM/KH calculation for the protein in the reactant and product sates to determine the 3D-density distribution $g_w(\mathbf{r};protein)$ of water around and inside protein, from which the radial distribution function $g_w(r;protein)$ is readily obtained.

4. Combining those steps above, one can calculate the friction $\zeta_{\alpha\alpha} = f_\alpha \zeta_{\alpha\alpha,bulk}$ of atoms in protein using Eqs. (23) and (24).

## V. Concluding remarks and perspective

A new theory to analyze the rate of chemical reactions in biomolecular solutions was proposed based on the generalized Langevin theory combined with the RISM/3D-RISM theory of molecular liquids. The theory employs the traditional picture of the transition state theory, developed by the greats in physical chemistry such as S. Arrhenius, H. Eyring, H. A. Kramers, and R. Marcus, in which a reaction proceeds by crossing over a barrier along a reaction coordinate, which is made by two free-energy surfaces representing the structural fluctuation of the two states of molecules, reactant and product. The free energy surfaces and its second derivatives with respect to the atomic coordinates, or Hessian, that characterize the structural fluctuation of biomolecules in solution, are obtained from the RISM/3D-RISM theory. It is suggested as a possible application of the theory to characterize the folding reaction of protein, for examples, whether it is a *barrier crossing* process between the two states or a *barrier-less* process.

    In the present paper, we just focused our attention on so-called "isomerization reaction" in which the reaction proceeds along an adiabatic surface. However, there are many chemical reactions of interest in biosystems, in which a transition between two adiabatic free-energy surfaces, or the non-adiabatic transition, becomes important for determining the reaction rate. The electron-transfer reactions taking place in proteins at photo-active center of plant are the most outstanding example of such reactions. In such reactions, the (classical) structural fluctuation and the electronic structure-change are highly conjugated. Three elements in the theoretical chemistry are required in addition to the theory on which

the present treatment is based: 1) the theory to evaluate the electronic structure of a molecule in solution; 2) the methodology to calculate the electronic structure of a macromolecule such as protein; and 3) the theory to describe the non-adiabatic transition between two adiabatic surfaces.

Concerning the electronic-structure of a molecule in solution, we have already developed a theory while ago, referred to as "RISM-SCF" which has been successfully applied to many problems related to chemical reactions, including the electron-transfer reactions.[41-45]

It is a non-trivial problem to calculate the electronic structure of a macromolecule, due mainly to its computational cost. However, the problem has become less serious lately in two reasons: the development of high-performance computer such as K-computer, and the development of models or/and methodologies in the quantum chemistry such as the ONIOM developed by Morokuma and his coworkers, and the fragment MO developed by Kitaura and his coworkers. [46, 47]

With respect to the non-adiabatic transition, a considerable progress has been made during the past few decades by H. Nakamura and his coworkers, further developing the idea proposed by Landau and Zener. [33, 48-50]

It will be a great challenge but a feasible task to put all those theories and methodologies together to calculate the rate of chemical reaction taking place in biomolecular systems based on the theory proposed here.

## Acknowledgement

The author is grateful to Prof. Nakamura for invaluable comments on the non-adiabatic transition between the two adiabatic surfaces in chemical reactions. He also thanks to Prof. Chong for providing the information concerning their study on the free energy landscape of protein in solution.